\documentclass[pre,twocolumn,bm,superscriptaddress,showpacs,showkeys,amsmath,amssymb,amsfonts]{revtex4-1}

\usepackage{ulem, fullpage, amsmath, amssymb, color, verbatim, graphicx, subfigure}

\def\s{\text{S}} \def\n{\text{N}} \def\t{\text{R}}

\def\tt{\text{RR}} \def\ts{\text{RS}} \def\tn{\text{RN}} \def\ss{\text{SS}}
\def\sn{\text{SN}} \def\nn{\text{NN}} \def\dt{\partial_t} 
\def\dtt{\partial_{\tilde{t}}}
\def\sN{\mathcal{N}}

\begin{document}

\title{Recruitment dynamics in adaptive social networks}

\author{Maxim S.~Shkarayev}
\affiliation{Applied Science Department, College of William \& Mary, Williamsburg, VA 23187}
\author{Ira B.~Schwartz}
\affiliation{Nonlinear Systems Dynamics Section, Plasma Physics Division, Code 6792, US Naval Research Laboratory, Washington, DC 20375}
\author{Leah B.~Shaw}
\affiliation{Applied Science Department, College of William \& Mary, Williamsburg, VA 23187}

%$^1$\textit{Applied Science Department, College of William \& Mary, Williamsburg, VA 23187}

\begin{abstract} We model recruitment in adaptive social networks in the presence 
of birth and death processes. Recruitment is characterized by nodes changing 
their status to that of the recruiting class as a result of contact with 
recruiting nodes. Only a susceptible subset of nodes can be recruited. The 
recruiting individuals may adapt their connections in order to improve 
recruitment capabilities, thus changing the network structure adaptively.  We 
derive a mean field theory to predict the dependence of the growth threshold of 
the recruiting class on the adaptation parameter. Furthermore, we investigate the 
effect of adaptation on the recruitment level, as well as on network topology.  
The theoretical predictions are compared with direct simulations of 
the full system. We identify two parameter regimes with qualitatively different 
bifurcation diagrams depending on whether nodes become susceptible frequently 
(multiple times in their lifetime) or rarely (much less than once per 
lifetime).\end{abstract}



\maketitle

\section{Introduction}

Any society contains individuals who are carriers of an ideology or fad (e.g., a 
religious or political party affiliation) that they desire to spread to the rest of the 
society. Thus, for a given ideology, a society can be partitioned into a set of people 
that represent the ideology and want to spread it, and the complement of this set. For 
example, the ideology could correspond to the views of a particular political party, 
with the party members desiring to recruit new members to improve their positions in the 
government. Other areas of recruitment have been proposed as mechanisms for fads which 
appear as a rapid rise above some threshold, as in music \cite{MeyerU10} ,management 
technologies \cite{BendorHW09}, economics \cite{JanssenJ01}, and even science 
\cite{Abrahamson09} . Slower recruitment based on social networks has been postulated in 
biology \cite{GrueterR11} and the spread of alcoholism \cite{Bender2006}. Slowing the 
rise of a fad or eliminating the spread of an ideology has been also been proposed 
through the control of critical nodes in a social network \cite{KuhlmanKMRR10}.

In today's world where collisions between ideologies can lead to radicalization of 
society, the problem of existence and formation of terrorist networks or insurgency 
movements becomes important. Recently, mathematical modeling of various radical groups, 
such as terrorist organizations, has been done to explore their structure and dynamics 
\cite{Gut12009,Gut2,Johnson2006,Cherif_10}. In addition to dynamical approaches that 
measure rates of attacks of radical groups \cite{Johnson2011}, operations research 
theories have also been applied to the formation of radical groups \cite{Caulkins2008}.  
Another class of works~\cite{Udwadia_06,Bentson,Butler2011} focuses on the recruitment 
dynamics of terrorists within a well-mixed population using compartmental models similar 
to those used for epidemic spread.  In \cite{Butler2011}, a systematic analysis of 
recruitment to Hezbollah has been done. Data from a political science discipline is used 
to generate a compartmental model of recruitment. Although the modeling is 
deterministic, it considers the various parameters which explicitly affect the success 
of recruitment to the radical's cause.  Finally, a number of recent publications discuss 
terrorist networks as optimal structures that optimize communication efficiency while 
balancing the secrecy of the networks~\cite{Lindelauf,Farley_03,Baccara_08}.

Not considered previously is the combination of the spread of radical ideas plus 
adaptive changes in social connections to improve recruitment success to the radical 
cause.  Some work in this direction is presented in the papers studying voter models in 
which individuals make connections to influence others' opinions 
\cite{Benczik2008,Benczik2009,Schmittmann2010}, as well as opinion dynamics models in 
which individuals are influenced by neighbors' opinions \cite{HolmeN06}.

We develop a simple model of a society in which some of the members belong to a class 
that tries to spread an ideology. The ideology spreads as a result of contact of 
recruiting members with nonrecruiting members. The recruiting members may improve their 
chances to recruit via network adaptation. The purpose of the adaptation proposed here 
is to improve the spread of the ideology, which is in contrast to network adaptation in 
epidemiological models~\cite{Gross2006b,Shaw2008a} where the purpose is avoidance of 
contact with the spreading members. In addition to adaptation, the connectivity within 
the society changes due to birth of new members and death of existing members. Using 
this model, we explore how the existence of stable recruiting classes depends on the 
adaptation and other parameters. We also consider the network topology of the recruiting 
class, as it may be important in accessing the quality of the communication channels in 
the resulting structure.  Section \ref{sec:model} presents the model and a system of 
mean field equations describing its dynamics.  Section \ref{sec:results} presents mean 
field analysis of the threshold for successful recruiting and its dependence on the 
network adaptation.  These results are compared with simulations of the full system, and 
the network geometry of the recruiting class is considered. Section 
\ref{sec:conclusions} concludes.

\section{Modeling the dynamics of recruitment}
\label{sec:model}
\subsection{Model}

We consider a social structure consisting of $M$ individuals represented by nodes in a 
network. An existing relation between any two individuals is represented by a link 
between the two nodes. New individuals join the society at a constant total rate $\mu$, 
and the individuals in the network can leave the network via death at rate $\delta$ per 
individual.  When new individuals join the system, they arrive with $\sigma$ links, 
which are connected to $\sigma$ randomly selected nodes in the population. When 
individuals die, their links are removed.

\begin{figure}[htb!]
\includegraphics[width=1.5in]{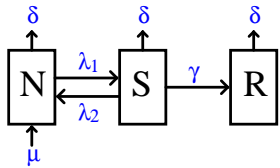}
\caption{Schematic representation of node fluxes in and out of the network due to
birth and death, described by the rates $\mu$ and $\delta$ respectively. Fluxes
between the three classes are described by the transition rates $\lambda_1$,
$\lambda_2$ and $\gamma$. Link rewiring (not shown here) takes place at rate $w$.
}\label{fig:scheme}
\end{figure}

Some members of the society, {\it recruiters} (also referred to as R-nodes), are 
carriers of an ideology that they try to spread to the individuals they come into 
contact with. The individuals that do not belong to the recruiting class are divided 
into two groups: those who are {\it susceptible} to the recruitment (S-nodes) and those 
who are {\it non-susceptible} (N-nodes). We assume that individuals can spontaneously 
change their state from non-susceptible to susceptible and vice versa (with rates 
$\lambda_1$ and $\lambda_2$ respectively). An individual in the recruiting class remains 
in that class until death~\cite{Udwadia_06}.

A susceptible individual joins the recruiting class at a rate that is proportional (with 
proportionality constant $\gamma$) to the number of contacts it has with the recruiting 
class. In order to improve their recruiting capabilities, the recruiting individuals can 
rewire their links with rate $w$, abandoning a connection to a nonsusceptible individual 
in favor of a connection to a susceptible one. As the links are rewired, the network 
topology changes based on the current states of its nodes. A schematic representation of 
the node dynamics is shown on Fig.~\ref{fig:scheme}.

We perform Monte Carlo simulations of the above system following the continuous time 
algorithm described in~\cite{Gillespie}. We start with an Erdos-Renyi random network, 
which then evolves according to the rules of birth, death, and rewiring. The results 
presented in the rest of the paper are for the systems that have reached a steady state.

\subsection{Mean field} To describe the dynamics of this system, we construct a mean 
field model for nodes and links as in~\cite{Gross2006b}. The time evolution of the nodes 
of each type is described by the following rate equations:
 \begin{subequations}\label{eq:nodes}
\begin{align}
\dt  \sN_{\n} &= \mu - \lambda_1 \sN_{\n}
+\lambda_2 \sN_\s- \delta \sN_{\n} \label{eq:node_N}\\
\dt  \sN_{\s} &= \lambda_1 \sN_{\n} -\lambda_2 \sN_{\s}- \gamma
\sN_\ts - \delta \sN_{\s} \label{eq:node_S}\\
\dt  \sN_{\t} &= \gamma \sN_\ts - \delta \sN_{\t}, \label{eq:node_R}
\end{align} 
 \end{subequations}
 Here the functions $\sN_{\n}\equiv\sN_{\n}(t)$, $\sN_{\s}\equiv\sN_{\s}(t)$, 
$\sN_{\t}\equiv\sN_{\t}(t)$ represent the number of nodes of each type. The process of 
recruitment is captured by the $\gamma \sN_{\ts}$ term, where the recruitment is shown 
to take place at a rate proportional to the number of links between the recruiting class 
and the susceptible class, $\sN_{\ts}$. 

In order to capture the rewiring process, we follow the evolution of the
number of different types of links present in the network:
\begin{subequations}\label{eq:links}
 \begin{align}
\begin{split}
 \dt \sN_\nn &= \lambda_2 \sN_\sn+
{\sigma \mu}\frac{\sN_{\n}}{\sN_\n+\sN_\s+\sN_\t}\\
& - 2(\lambda_1+\delta) \sN_\nn
\end{split}\\
\begin{split}
 \dt \sN_\sn &=
{\sigma \mu}\frac{\sN_{\s}}{\sN_\n+\sN_\s+\sN_\t}-
\gamma \sN_{\text{NSR}}+2\lambda_2 \sN_\ss \\
&-(\lambda_1 +\lambda_2+2\delta)\sN_\sn+ 2\lambda_1 \sN_\nn
\end{split}\\
\begin{split}
 \dt \sN_\ss &=
- \gamma \sN_{\text{SSR}}+\lambda_1 \sN_\sn \\
&-2(\lambda_2+\delta) \sN_\ss
\end{split}\\
\begin{split}
\dt \sN_\tn &= \gamma \sN_{\text{NSR}}+{\sigma \mu}\frac{\sN_{\t}}{\sN_\n+\sN_\s+\sN_\t}\\
&-(\lambda_1 +2\delta+w)\sN_\tn +\lambda_2 \sN_\ts
\end{split}\\
\begin{split}
 \dt \sN_\ts &=- \gamma \sN_{\text{RSR}}+ \gamma \sN_{\text{SSR}}\\
&-(\lambda_2+2\delta) \sN_\ts +(\lambda_1+w) \sN_\tn
\end{split}\\
\begin{split}
 \dt \sN_\tt &= \gamma\sN_{\text{RSR}}  -{2\delta \sN_\tt}.
\end{split}\label{eq:link_RR}
\end{align}
\end{subequations}
where the terms $\sN_{\text{xy}}$ correspond to the number of links connecting nodes 
from classes x and y, e.g., $\sN_{\text{\nn}}$ corresponds to the number of NN links. 
The terms proportional to $\sigma \mu$ correspond to the influx of edges due to the 
birth of new nodes, where the probability for the new node to attach itself to a node 
from class X is proportional to the number of nodes in that class. The third order 
terms, $\sN_{\text{NSR}}$, $\sN_{\text{SSR}}$ and $\sN_{\text{RSR}}$, describe the 
formation of triples of nodes with an S-node at the center, and at least one of the 
edges terminating at an R-node. These terms describe the rate at which NS-, SS- and 
RS-links become NR-, SR- and RR- links respectively, due to the interaction of the 
central S-node with its neighboring R-node. Note that our definition of RSR triples 
includes {\it degenerate} triples, i.e., RSR triples where both of the R-nodes 
correspond to a single R-node, by analogy with degenerate triangles.

The resulting system of equations is not closed, as it contains higher order terms. 
Following earlier works in epidemiology~\cite{Keeling1997}, we introduce the closure 
based on the assumption of homogeneous distribution of $\ts$-links:
 \begin{subequations}\label{eq:closure}
\begin{align}
\sN_{\text{NSR}}&=\frac{\sN_\ts}{\sN_\s}\frac{\sN_\sn}{\sN_\s} \sN_\s\\
\sN_{\text{SSR}}&=\frac{\sN_\ts}{\sN_\s}\frac{2\sN_\ss}{\sN_\s} \sN_\s\\
\sN_{\text{RSR}}&=\left(\left(\frac{\sN_\ts}{\sN_\s}\right)^2+\frac{\sN_\ts}{\sN_\s}\right)\sN_\ts.
\end{align}
 \end{subequations}
We thus obtain a system of nine mean field equations, which we analyze.

\section{Results} \label{sec:results} 
\subsection{Mean field recruiting threshold}\label{subsec:res_1}
 We first consider the bifurcation point of the recruitment model where the zero recruit 
(trivial) steady state becomes unstable, which we call the {\it recruiting threshold}. 
This is a transcritical bifurcation point, which is analogous to the epidemic onset in 
epidemic spreading models. It can be found analytically as a function of parameters for 
the mean field model, and certain asymptotic limits have a simple form.

We nondimensionalize Eqns.~(\ref{eq:nodes}) and~(\ref{eq:links}) to simplify the 
analysis.  We introduce the dimensionless time variable $\tilde{t} \equiv \delta t$, 
where time is rescaled by the average node lifetime $\delta^{-1}$. Further, the node 
variables are rescaled by the expected population size at steady state, $\mu/\delta$, 
while the link variables are rescaled by the expected number of links at steady state, 
$(\mu/\delta)(\sigma/2)$. Let ${\bf x}\equiv[\sN_\n, \sN_\s, 
\sN_\t, \sN_\nn2\sigma^{-1}, \sN_\sn2\sigma^{-1}, \sN_\ss2\sigma^{-1},\\ 
\sN_\tn2\sigma^{-1}, \sN_\ts2\sigma^{-1}, \sN_\tt2\sigma^{-1} ]\delta \mu^{-1}$ denote 
the 9-dimensional vector of rescaled node and link state variables. Note that in the 
steady state the rescaled node variables sum to 1, and, therefore, in steady state they 
correspond to the probability for a node of a given type to exist in the system. 
Similarly, the rescaled link variables sum to 1, and correspond to the probability for 
an edge chosen at random to be of a given type, e.g., at steady state $x_{8}$ 
corresponds to the probability for a randomly chosen edge to be an RS link.

We introduce the following
rescaled parameters:
\begin{subequations}
\begin{align}
\Lambda_1& \equiv \lambda_1 \delta^{-1}\\
\Lambda_2& \equiv \lambda _2 \delta^{-1}\\
\Gamma& \equiv(\gamma \sigma/2)\delta^{-1}\\
W& \equiv w \delta^{-1}
\end{align}
\end{subequations}
We can now write the dimensionless equations of motion as
\begin{align}
\dot{\bf x}={\bf F}({\bf x};\bar{\Lambda}).\label{eq:EOM}
\end{align}
where $\bar\Lambda\equiv[\Lambda_1,\Lambda_2,W,\Gamma,\sigma]$ is a vector of
all the system parameters.  (Recall that $\sigma$ is a dimensionless integer.)
The full system of dimensionless equations is given in Eqs.~(\ref{eq:dimensionless}).

We can now find the trivial steady state solution, where the number of $\t$ nodes is 
zero. This restricts the state to be of the form ${\bf 
x}_0=[x_{0,1},x_{0,2},0,x_{0,4},x_{0,5},x_{0,6},0,0,0 ]$, where the number of links 
involving $\t$ nodes is zero as well. This guarantees that $F_{i}({\bf x}_{0},\bar 
\Lambda)=0$ for $i=3,7,8,9$. Since this subset of equations represents an invariant 
manifold, we concentrate on solving the equations for the rest of the 5 
components, $x_{0,1}$, $x_{0,2}$, $x_{0,4}$, $x_{0,5}$ and $x_{0,6}$.

The first observation is that the $x_{1}$ and $x_{2}$ equations are solvable when
the number of recruiting nodes is zero, and they yield steady state values of
\begin{subequations}
\begin{align}
x_{0,1}&= (\Lambda_{2}+1)D^{-1}, \\
x_{0,2}&= \Lambda_1 D^{-1},\label{eq:x1x2 solution}
\end{align}
where $D \equiv  \Lambda_{1}+\Lambda_2+1$.
The nonzero link variables may be expressed in terms of $x_{0,1},x_{0,2}$,
and they are given by the following:
\begin{align}
x_{0,4} & = (\Lambda_{2}+1)^2  D^{-2}, \\
x_{0,5} & =  2\Lambda_{1}(\Lambda_{2}+1) D^{-2},\\
x_{0,6} & =  \Lambda_{1}^{2} D^{-2}.
\label{eq:x4x5x6 solution}
\end{align}
\end{subequations}

Now that we have the full trivial solution, we can examine its stability. In
order to do this, we linearize the vector field about ${\bf x_{0}}$ and examine
where it has a
one dimensional null space. That is, at the bifurcation point, there is only one
real eigenvalue passing through zero. This is equivalent to examining where the
determinant of the Jacobian vanishes; i.e., we compute those parameters where
\begin{equation}
\det(\mathcal{D}_{{\bf x}}{\bf F}({\bf x}_{0},\bar \Lambda))=0. \label{eq:determ of F}
\end{equation}
The following relation describes the location of the bifurcation:
 \begin{align}
 W=-\Lambda_1 \left[ 1+\frac{2\Gamma} {[\Gamma
(2-\sigma^{-1})-1](\Lambda_1+\Lambda_2+1)}\right] \nonumber\\
+\Lambda_2 \frac{1}{\Gamma (2-\sigma^{-1})-1} +\frac{2(\Gamma \sigma^{-1}+1)}{\Gamma
(2-\sigma^{-1})-1}
 \label{eq:W}
 \end{align}

We next examine the limits of Eq.~(\ref{eq:W}) when either the recruitment rate
$\Gamma$ or rewiring rate $W$ is large.   The
minimum amount of recruitment required to maintain nonzero values of
recruited population, when $W$ approaches infinity, can be found by setting the least common denominator of
Eq.~(\ref{eq:W}) to zero, which yields a simple expression for $\Gamma$:
 \begin{align}
\Gamma=\frac{1}{2-\sigma^{-1}} \label{eq:asympt_vert}
 \end{align}
If the recruitment rate is below this value, additional rewiring is not sufficient to enable spreading of the recruiting class.

Similarly, we
find the smallest amount of rewiring required for existence of a nontrivial
steady state solution, conditioned on our ability to control $\Gamma$, with all
the other parameters fixed:
 \begin{align*} \lim_{\Gamma \to \infty} W
&=-\Lambda_1 \left[ 1+\frac{2} {(2-\sigma^{-1})(\Lambda_1+\Lambda_2+1)}\right]+\\
&+2\frac{\sigma^{-1}}{2-\sigma^{-1}}
 \end{align*}
For lower rewiring rates, the recruiting class cannot spread even if the recruiting rate is large.  The value of this asymptote is greatest, meaning rewiring is most necessary, when $\Lambda_1$ approaches zero. In
this case few non-susceptible nodes become susceptible to recruiting, and the necessary rewiring value approaches
 \begin{align}
W=\frac{2\sigma^{-1}}{2-\sigma^{-1}} \label{eq:asympt_horiz}
 \end{align}

\begin{figure}[htb!]
\subfigure[]{\includegraphics{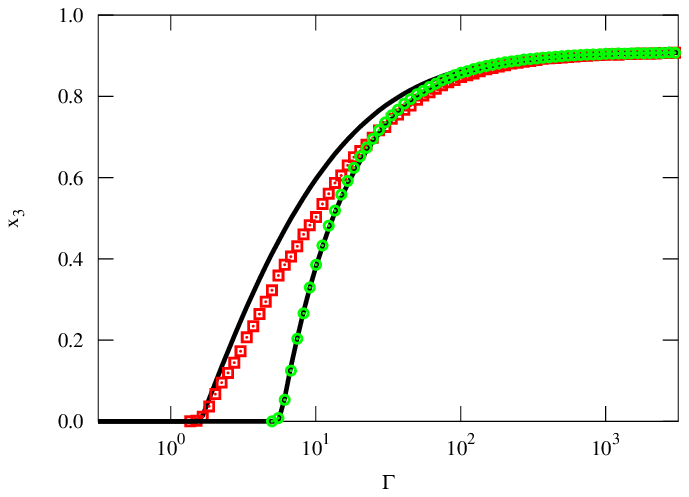}\label{fig:TX1000}}
\subfigure[]{\includegraphics{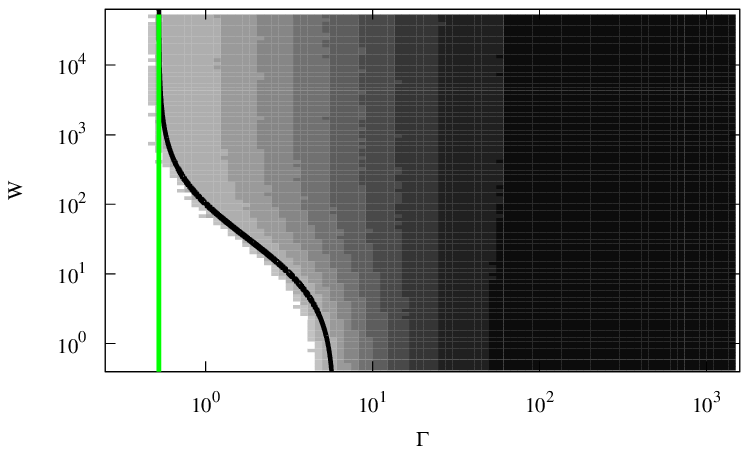}\label{fig:T1000}}
\caption{Direct numerical simulation of a stochastic network system with $\Lambda_1 \gg 
1$. In Fig.~\ref{fig:TX1000} we compare the result of the simulation (symbols) to the 
mean field solution given by Eq.~(\ref{eq:x3_exact}) (solid curves). Square (red 
online): strong rewiring, $W\approx 40$. Circle (green online): weak rewiring, $W\approx 
0.4$. Fig.~\ref{fig:T1000} shows the density plot representing the dependence of the 
fraction of recruited nodes in the population in statistical steady state on rescaled 
rates of rewiring, $W$, and recruiting, $\Gamma$. Vertical solid line (green online) 
corresponds to the minimum recruitment rate required to support a nontrivial solution, 
as given by Eq.~(\ref{eq:asympt_vert}). The solid curve (black online)
corresponds 
to the recruiting threshold as predicted by mean field in Eq.~(\ref{eq:W}). 
The simulations are performed on a system with $\mu=10^5$, $\delta=1$, $\sigma=10$, 
$\lambda_1=10$, $\lambda_2=100$. The time averaged values are computed once the system 
reaches steady state regimes. (In all of the figures in this paper this corresponded to 
at least $10 \delta^{-1}$ units of time, in other words about 10 generations of nodes 
have died before we assume the system to be at the steady state.)
}\label{fig:simT1000}
 \end{figure}

\begin{figure}[htb!]
\subfigure[]{\includegraphics{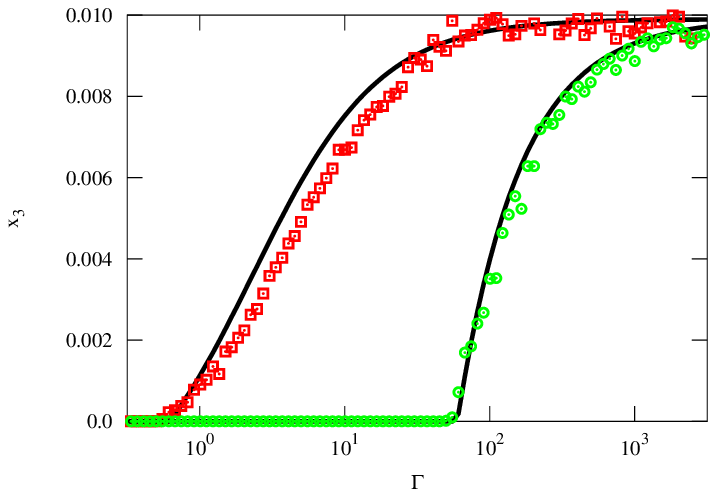}\label{fig:TX0001}}
\subfigure[]{\includegraphics{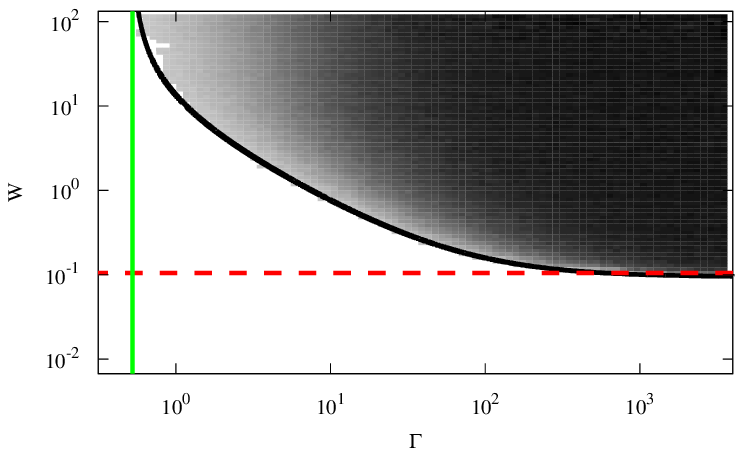}\label{fig:T0001}}
\caption{Direct numerical simulation of a stochastic network system with 
$\Lambda_1 \ll 1$. In Fig.~\ref{fig:TX0001} we compare the result of the simulation to 
the mean field solution given by Eq.~(\ref{eq:x3_exact}). Square (red online): strong 
rewiring, $W\approx70$. Circle (green online): weak rewiring, $W\approx0.2$. 
Fig.~\ref{fig:T0001} shows the density plot representing the dependence of the fraction 
of recruited nodes in the population in statistical steady state on rescaled rates of 
rewiring, $W$, and recruiting, $\Gamma$. Vertical solid line (green online) corresponds 
to the minimum recruitment rate required to support nontrivial solution, as given by 
Eq.~(\ref{eq:asympt_vert}). Horizontal dashed line (red online)
corresponds to the rewiring rate that guarantees existence of a nontrivial solution in 
the limit of large recruitment rate 
(Eq.~\ref{eq:asympt_horiz}).
The solid curve corresponds to 
the recruiting threshold as predicted by mean field in Eq.~(\ref{eq:W}).  
The simulations are performed on a system with $\mu=10^5$, 
$\delta=1$, $\sigma=10$, $\lambda_1=0.01$, $\lambda_2=10$. }\label{fig:simT0001}
\label{fig:simT0001}
\end{figure}

\subsection{Comparison of mean field with simulations}

 We next compare the mean field predictions for the spread of recruiters with 
simulations of the full stochastic network system. Thus, we compare the average size of 
the recruited portion of the population in statistical steady state to the solution of 
the mean field equations at steady state, which we solve exactly in the 
Appendix~\ref{app:exact}.  We assume that the parameters $\Lambda_1$ and $\Lambda_2$ 
(rates for gaining and losing susceptibility) and $\sigma$ (determines average degree) 
depend on details of the society.  On the other hand, we assume that the parameters 
$\Gamma$ and $W$ can be controlled by the recruiters, i.e., the recruiters may choose to 
be more or less aggressive in their recruitment, as well as in how quickly they rewire 
their links to susceptible members of society. Therefore, we investigate the recruiting 
effectiveness for a given choice of $\Lambda_1$, $\Lambda_2$ and $\sigma$, while varying 
$\Gamma$ and $W$.

We distinguish two parameter regimes, corresponding to two qualitatively different 
behaviors of the system: small values of $\Lambda_1$ ($\Lambda_1 \ll 1$) and large 
values of $\Lambda_1$ ($\Lambda_1 \gg 1$).  The small $\Lambda_1$ regime corresponds to 
the case where nodes become susceptible to recruiting only rarely, on average much less 
than once in their lifetime (where the average lifetime of a node is $\delta^{-1}$ in 
the original time units and $1$ in the dimensionless $\tilde{t}$ units). The large 
$\Lambda_1$ regime corresponds to the case where, on average, nodes become susceptible 
to recruitment at least once in their lifetime.

In Figs.~\ref{fig:simT1000} and~\ref{fig:simT0001} we show the results of
simulating the system in these two regimes. The density plots in
Figs.~\ref{fig:T1000} and~\ref{fig:T0001} show the fraction of recruited nodes as
a function of recruiting rate $\Gamma$ and rewiring rate $W$ in networks with
$\Lambda_1 \gg 1$ and $\Lambda_1\ll 1$ respectively. The black curves in the two
figures represent the location of the recruiting threshold as derived from the
mean field equations and given by Eq.~(\ref{eq:W}). Here mean field allows us
to accurately predict the onset of the stable nontrivial solution. In
Figs.~\ref{fig:TX1000} and~\ref{fig:TX0001} we compare the simulation results to
the mean field predictions. Even though there is some discrepancy between the
direct simulations and the mean field near the bifurcation, the recruiting
threshold and the asymptotic behavior for large $\Gamma$ show excellent agreement
with the simulations.

We observe in simulations that the steady state trivial solution undergoes a forward 
transcritical bifurcation in the recruiting rate $\Gamma$ for all values of $W$ for 
which a nontrivial solution exists. This is in contrast with epidemiological models 
where the purpose of the rewiring is avoidance of the nodes spreading 
infection~\cite{Shaw2008a,Gross2006b}, which can undergo a backward transcritical 
bifurcation and exhibit bistability. Note that, unlike those models, the purpose of the 
rewiring in the recruitment model is attraction of the susceptible population by the 
recruiters.

We would like to draw the reader's attention to an important difference in the 
system behavior in the two regimes. In the regime where $\Lambda_1 \gg 1$, the 
nontrivial solution exists for all values of $W$ as long as $\Gamma$ is large 
enough. On the other hand, in the regime where 
$\Lambda_1\ll 1$, the nontrivial solution may fail to exist for any value of 
$\Gamma$ unless rewiring is aggressive enough. In Fig.~\ref{fig:T0001} there is a 
range of rewiring rates $W$ for which only trivial solutions exist. The 
horizontal dashed line, given by Eq.~(\ref{eq:asympt_horiz}), indicates the mean 
field level of rewiring that is {\it sufficient} for the nontrivial solution 
to exist for rapid recruiting (large $\Gamma$). Also, we can see that the 
rewiring level that is {\it necessary} for the emergence of the nontrivial 
solution can be very close to the predicted sufficient level of 
Eq.~(\ref{eq:asympt_horiz}), for small values of $\Lambda_1$.
\begin{figure}[htb!]
\includegraphics{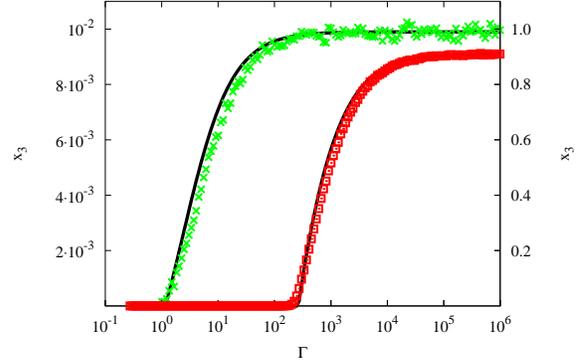}
\caption{
Comparing systems with different values of $\Lambda_1$, but same ratio of
$\Lambda_1/(1+\Lambda_2)$.
Squares (red online), right axis: $\Lambda_1=10$, $\Lambda_2=10^4$.
Crosses (green online), left axis: $\Lambda_1=0.01$, $\Lambda_2=9.001$.
Both of the simulations were performed with the following
parameters: $w=10$, $\mu=10^5$, $\delta=1$, $\sigma=10$.
}\label{fig:l1l2}
\end{figure}

Another important difference between the two regimes of $\Lambda_1$ 
values is seen when we compare two systems with different values of 
$\Lambda_1$ and $\Lambda_2$ while the ratio $\Lambda_1/(1+\Lambda_2)$ 
is kept fixed. Note that in the absence of recruitment, such systems 
have identical fractions of susceptible nodes. In other words, in the 
presence of recruitment, the pool of individuals available for 
recruitment would appear to be the same. However, as we can see from 
Fig.~\ref{fig:l1l2}, there is a significant difference in the size of 
the recruited population as well as the recruiting threshold. The 
difference appears to be caused by the difference in the dynamics in 
the two regimes. Thus, for the large values of $\Lambda_1$ and 
$\Lambda_2$ a node can become susceptible several times during its 
life, while as the value of $\Lambda_1$ decreases, only some nodes will 
ever become susceptible, with low likelihood of doing so more than 
once.

\begin{figure}[htb!]
\includegraphics{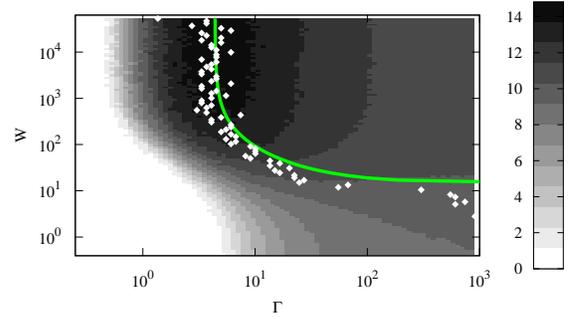}
\caption{ Direct numerical simulation. Density plot representing the dependence of the 
fraction of the mean node degree in the R-subnetwork in steady state, 
$2\sN_\tt/\sN_\t$, on rescaled rates of rewiring, $W$, and recruiting, $\Gamma$. Solid 
line (green online) indicates the value of $\Gamma$ where the degree is at a maximum for 
a given value of $W$, found using analytic expression in Eq.~(\ref{eq:gamma_m}). The 
white diamonds correspond to the location of the maximum in simulations for a given 
value of $W$. The simulations are performed with the following parameters: $\mu=10^5$, 
$\delta=1$, $\sigma=10$, $\lambda_1=10$, $\lambda_2=100$.
}\label{fig:TT1000}
\end{figure}

\subsection{Recruited subnetwork geometry}

We now investigate the structure of the portion of the network (later referred to as the 
R-subnetwork) consisting only of R-nodes and links between them. In particular, we are 
interested in the mean degree of its nodes when the system reaches a steady state. The 
mean field approach, in addition to predicting the fraction of recruited nodes (which 
corresponds to the size of the R-subnetwork), also provides information about topology 
of the subnetwork in the form of mean degree of the nodes. The mean degree of a node in 
the subnetwork serves as a low order description of how well the nodes in the network 
are connected, which may be important for communication within the established 
subnetwork. The density plot in Fig.~\ref{fig:TT1000} shows the dependence of the mean 
degree on $W$ and $\Gamma$.  The nonmonotonic behavior as a function of $\Gamma$ for 
fixed $W$ is predicted by the mean field model. The solid line corresponds to the 
analytical prediction of the maximum's location as given by Eq.~(\ref{eq:gamma_m}) 
derived in Appendix~\ref{app:extremum}. Note that the analytic description of the 
maximum's location would allow the recruiting class to optimize connectivity within the 
subnetwork, if that happens to be an important goal for that class. The mean field 
approximation fails to capture the nonmonotonic behavior of the mean degree in the limit 
where $\Lambda_1 \ll 1$, and we leave this issue to a future study.

\section{Conclusions and Discussion}\label{sec:conclusions}
 We develop a toy model that describes recruitment of new members by an interested class 
within a society. The members of the recruiting class can improve their recruiting 
capabilities by via adaptation. Thus, they may choose to abandon their relations with 
those members of society that are not prone to recruitment. This model assumes that once 
a node joins the recruiting class, it itself becomes a recruiter and it remains a member 
of this class until death.  In contrast to avoidance rewiring used to reduce infection 
spreading in epidemic models~\cite{Gross2006b,Shaw2008a}, network adaptation in our 
model promotes spreading.  Additionally, the population is open with birth and death 
modeled explicitly, while most previous adaptive network models have been of closed 
populations (e.g., 
\cite{Gross2006b,Shaw2008a,Gross2008a,Benczik2008,Benczik2009,Schmittmann2010,HolmeN06}).

In this paper, we develop and analyze a mean field description of our model. Thus, we 
are able to accurately predict the onset of the stable nontrivial solution of the system 
at steady state. Furthermore, we are able to accurately predict the size of the 
recruited class for a given set of system parameters, as well as the mean degree of the 
subnetwork formed by the recruited members. We compare the predictions made by the mean 
field description with the direct simulations of our model. We generally observe a good 
agreement between the model and its mean field approximation.

Analyzing the mean field model, we find two parameter regimes with very distinct 
qualitative behavior. Thus, we show that in the society where the particular ideology is 
unpopular (perhaps corresponding to radical ideology) and individuals rarely become 
susceptible to it, adaptation is necessary in order to observe stable nonzero levels of 
the recruiting class. On the other hand, if the idea is sufficiently popular (e.g., 
recruitment into a moderate political party), the adaptation improves the recruiting 
capabilities and may affect the ultimate topology of the recruited, but adaptation is 
not a necessary condition for the existence of a nonzero stable solution. Furthermore, 
we speculate that if the model were changed to describe a society with two competing 
recruiting classes (think two party system), the adaptation may be a mechanism by which 
one competing class gets an edge over the other class.

As evidenced by the results presented in Figs.~\ref{fig:TX1000} and~\ref{fig:TX0001}, 
the mean field approximation has some inaccuracies in the parameter regime between the 
bifurcation point and the asymptotic saturation.  We predict that this inaccuracy is due 
to errors in the homogeneous closure assumption (Eqs.~\ref{eq:closure}).  In a 
subsequent work, we develop and analyze a new closure of the mean field equations that 
more accurately describes the full system.

\begin{acknowledgments}

MSS and LBS were supported by the Army Research Office, Air Force Office of Scientific 
Research, and by Award Number R01GM090204 from the National Institute Of General Medical 
Sciences.  IBS was supported by the Office of Naval Research, the Air Force Office of 
Scientific Research, and the National Institutes of Health. The content is solely the 
responsibility of the authors and does not necessarily represent the official views of 
the National Institute Of General Medical Sciences or the National Institutes of Health.

\end{acknowledgments}

\appendix
\section{Exact solution}\label{app:exact}

In this appendix, we derive the exact solution to the system of equations in 
Eqs.~(\ref{eq:nodes}) and~(\ref{eq:links}) when the system is at the nontrivial steady 
state, i.e., when the left hand side of the equations is zero and the size of recruiting 
class is nonzero. We begin by deriving the dimensionless equations, as defined in 
section~\ref{subsec:res_1}:
 \begin{subequations}\label{eq:dimensionless}
\begin{align}
\dtt  x_1 &= 1 - \Lambda_1 x_1 +\Lambda_2 x_2-  x_1\\
\dtt  x_2 &= \Lambda_1 x_1 -\Lambda_2 x_2- \Gamma x_8 -  x_2\\
\dtt  x_3 &= \Gamma x_8 -  x_3\\
\dtt x_4 &= \Lambda_2 x_5+2\frac{x_1}{x_1+x_2+x_3} - 2(\Lambda_1+1) x_4\\
\dtt x_5 &= 2\frac{x_2}{x_1+x_2+x_3}- \Gamma \frac{x_5 x_8}{x_2}+2\Lambda_2 x_6
\nonumber\\
&-(\Lambda_1 +\Lambda_2+2)x_5+ 2\Lambda_1 x_4 \\
\dtt x_6 &= - 2\Gamma \frac{x_8 x_6}{x_2}+\Lambda_1 x_5 -2(\Lambda_2+1) x_6\\
\dtt x_7 &=  \Gamma \frac{x_5 x_8}{x_2}+2\frac{x_3}{x_1+x_2+x_3} \nonumber\\
&-(\Lambda_1 +2+W)x_7 +\Lambda_2 x_8\\
\dtt x_8 &= - \Gamma \left(\frac{x_8^2}{x_2}+2\sigma^{-1}x_8\right)
+ 2\Gamma \frac{x_6 x_8}{x_2}\nonumber\\
&-(\Lambda_2+2) x_8 +(\Lambda_1+W) x_7 \\
\dtt x_9 &= \Gamma \left(\frac{x_8^2}{x_2}+2\sigma^{-1}x_8\right) -{2 x_9}.
\end{align}
 \end{subequations}
At steady state, the left hand side of the above equations is zero. We proceed in
our derivation by dividing all equations in the steady state
by $x_2$ and introducing a new variable $z_i\equiv
x_i/x_2$, obtaining the following system of equations:
\begin{subequations}
\begin{align}
0 &= 1/x_2 - \Lambda_1 z_1 +\Lambda_2-  z_1 \label{eq:z1}\\
0 &= \Lambda_1 z_1 -\Lambda_2 - \Gamma z_8 - 1 \label{eq:z2}\\
0 &= \Gamma z_8 -  z_3 \label{eq:z3}\\
0 &= \Lambda_2 z_5+2z_1 - 2(\Lambda_1+1) z_4\label{eq:z4}\\
0 &= 2- \Gamma z_5 z_8+2\Lambda_2 z_6 -(\Lambda_1 +\Lambda_2+2)z_5\nonumber\\
&+ 2\Lambda_1 z_4 \label{eq:z5}\\
0 &= - 2\Gamma z_8 z_6+\Lambda_1 z_5 -2(\Lambda_2+1) z_6 \label{eq:z6}\\
0 &=  \Gamma z_5 z_8+2z_3-(\Lambda_1 +2+W)z_7 +\Lambda_2 z_8 \label{eq:z7}\\
0 &= - \Gamma \left(z_8^2+2\sigma^{-1}z_8\right)+ 2\Gamma z_6 z_8 \nonumber\\
&-(\Lambda_2+2) z_8 +(\Lambda_1+W) z_7 \label{eq:z8}\\
0 &= \Gamma \left(z_8^2+2\sigma^{-1}z_8\right) -{2 z_9} \label{eq:z9}.
\end{align}
\end{subequations}
We have used the fact that in the steady state $x_1+x_2+x_3=1$, as can be shown
by adding together Eqs.~(\ref{eq:node_N})-(\ref{eq:node_R}).
In the rest of the derivation we will find a closed equation for $z_8$ and
express the other $z_i$'s in terms of $z_8$.

We can immediately express $z_1$, $z_3$, and $z_9$ in terms of $z_8$ by solving 
Eqs.~(\ref{eq:z2}), (\ref{eq:z3}) and~(\ref{eq:z9}) respectively:
\begin{align}
z_1 &= \Lambda_1^{-1}(1+\Lambda_2 + \Gamma z_8 )\label{eq:z1_sol}\\
z_3 &= \Gamma z_8 \label{eq:z3_sol}\\
z_9 &= (\Gamma/2) \left(z_8^2+\sigma^{-1}z_8\right).
\end{align}
We substitute Eq.~(\ref{eq:z1_sol}) into Eq.~(\ref{eq:z4}), and
Eq.~(\ref{eq:z3_sol}) into Eq.~(\ref{eq:z7}). The resulting two equations,
together with Eqs.~(\ref{eq:z5}),~(\ref{eq:z6}) and~(\ref{eq:z8}), form a closed
system
of five equations with five unknowns $z_4$-$z_8$.

We solve Eq.~(\ref{eq:z6}) for $z_5$ in terms of $z_6$ and $z_8$:
 \begin{align}
z_5=2 \Lambda_1^{-1}(1+\Lambda_2+\Gamma z_8)z_6 \label{eq:z5_sol},
 \end{align}
 and substitute the above result into Eq.~(\ref{eq:z4}) and~(\ref{eq:z7}), to solve for 
$z_4$ and $z_7$ in terms of $z_6$ and $z_8$ as follows:
 \begin{align}
z_4 &=  [(\Lambda_1+1)\Lambda_1 ]^{-1}[1+\Lambda_2+\Gamma z_8][1+\Lambda_2 z_6]\label{eq:z4_sol}\\
z_7&=[ \Gamma  \Lambda_1^{-1}(2\Gamma z_8+2(\Lambda_2+1))z_6+2\Gamma
+\Lambda_2] \times\nonumber\\
& \times(\Lambda_1 +2+W)^{-1}z_8.\label{eq:z7_sol}
 \end{align}
 Substituting the expressions for $z_4$ and $z_5$ from Eq.~(\ref{eq:z5_sol}) 
and~(\ref{eq:z4_sol}) into Eq.~(\ref{eq:z5}), and solving for $z_6$ we obtain
 \begin{align}
z_6=\frac{\Lambda_1}{\Gamma(\Lambda_1+1)z_8+(\Lambda_1+\Lambda_2+1)}\label{eq:z6_sol},
 \end{align}
 Finally, substituting results of Eqs.~(\ref{eq:z6_sol}) and~(\ref{eq:z7_sol}) into 
Eq.~(\ref{eq:z8}) we obtain an equation for $z_8$ in a closed form:
 \begin{align}
\frac{
[(a_1 \Gamma) z_8^2+ (a_2 \Gamma +a_3)z_8+ (a_4 \Gamma^{-1}+a_5)]z_8}
{[\Gamma(\Lambda_1+1)z_8+(\Lambda_1+\Lambda_2+1)]}=0 \label{eq:z8_sol}
 \end{align}
where
 \begin{align}
a_1 & \equiv (\Lambda_1+1)(\Lambda_1+W+2)\\
a_2 &\equiv
2\sigma^{-1}(\Lambda_1+1)(\Lambda_1+W+2)\nonumber\\
&-2(\Lambda_1+2)(W+\Lambda_1)\\
a_3 & \equiv 3(\Lambda_1+1)(\Lambda_1+W+2+\Lambda_2)+\Lambda_2(W+1)\\
a_4 & \equiv 2(\Lambda_1+\Lambda_2+1)(\Lambda_1+W+2+\Lambda_2)\\
a_5 & \equiv(\Lambda_1+\Lambda_2+1)[2(\sigma^{-1}-2)(W+\Lambda_1)\nonumber\\
&+4(\sigma^{-1}-\Lambda_1)].
 \end{align}
 Note that the physically relevant solutions are positive, and therefore finding
a nontrivial solution of $z_8$ is a simple matter of solving a quadratic equation:
 \begin{align}
(a_1 \Gamma) z_8^2+ (a_2 \Gamma +a_3)z_8+ (a_4 \Gamma^{-1}+a_5)=0.\label{eq:quad}
 \end{align}

Solving Eq.~(\ref{eq:z1}) for $x_2$, we can now express $x_2$ in terms of the newly 
found $z_8$:
 \begin{align}
x_2=\Lambda_1[(\Lambda_1+1)\Gamma z_8+(\Lambda_1+\Lambda_2+1)]^{-1}.
 \end{align}
 The rest of the original variables can be found using $x_i=x_2 z_i$. Thus, for example, 
$x_3$ is
 \begin{align}\label{eq:x3_exact}
x_3=x_2 z_3 = \Gamma \Lambda_1 z_8 [(1+ \Lambda_1)\Gamma
z_8+(\Lambda_1+\Lambda_2+1)]^{-1}.
 \end{align}

\section{Extremum of the mean degree in R-subnetwork}\label{app:extremum}
 In steady state, the mean degree of the nodes within the R-subnetwork 
is found by taking a ratio of twice the number of $\tt$-links to the number of 
$\t$-nodes in the subnetwork: 
 \begin{align} \langle k \rangle = \frac{2 
\sN_\tt}{\sN_\t}= \frac{\sN_\ts}{\sN_\s}+1=\frac{\sigma}{2} z_8+1.
\end{align}
 where the values of $\sN_\tt$ and $\sN_\t$ are obtained by solving 
Eqs.~(\ref{eq:node_R}) and~(\ref{eq:link_RR}) in steady state. In this appendix we 
determine the value of $\Gamma$, $\Gamma_m$, that for a given rewiring rate will 
maximize the degree in the resulting R-subnetwork.  We do this by maximizing $z_8$.

Differentiating Eq.~(\ref{eq:quad}) with respect to $\Gamma$ and evaluating the 
resulting equation at $\Gamma_m$, the value where extremum is attained, we obtain
 \begin{align}
a_1 (z_8)_m^2+a_2 (z_8)_m -\Gamma_m^{-2}a_4=0.\label{eq:quad_der}
 \end{align}
 Note that the derivative of $z_8$ with respect to $\Gamma$ evaluated at $\Gamma_m$ is 
equal to zero because $z_8$ is an extremum there, and the $a_i$ are independent of 
$\Gamma$. Multiplying the above equation by $\Gamma$ and subtracting it from 
Eq.~(\ref{eq:quad}) evaluated at $\Gamma_m$ allows us to solve for $(z_8)_m$:
 \begin{align}
(z_8)_m = -\frac{2a_4 + \Gamma a_5}{\Gamma a_3},
 \end{align}

Substituting the value of $(z_8)_m$ into Eq.~(\ref{eq:quad_der}) and solving for
$\Gamma_m$ we obtain
 \begin{align}
 &\Gamma_m=[a_2 a_3 a_4-2 a_1 a_4 a_5 \nonumber\\ &+(a_2^2 a_3^2 a_4^2+a_1 
a_5^2 a_4 a_3^2-a_2 a_3^3 a_5 a_4)^{1/2}]\label{eq:gamma_m}\\ &/[a_5 (a_1a_5-a_3 
a_2)],\nonumber
 \end{align}
the location of the maximum for the given rewiring rate.

\bibliographystyle{apsrev}

%\bibliography{bibliography_new,bibliography_new2,more_refs,fads_merged}
%\bibliography{Shkarbib}

\end{document}